# Upper Atmospheres and Ionospheres of Planets and Satellites

Antonio García Muñoz, Tommi T. Koskinen, and Panayotis Lavvas

**Abstract**

The upper atmospheres of the planets and their satellites are more directly exposed to sunlight and solar-wind particles than the surface or the deeper atmospheric layers. At the altitudes where the associated energy is deposited, the atmospheres may become ionized and are referred to as ionospheres. The details of the photon and particle interactions with the upper atmosphere depend strongly on whether the object has an intrinsic magnetic field that may channel the precipitating particles into the atmosphere or drive the atmospheric gas out to space. Important implications of these interactions include atmospheric loss over diverse timescales, photochemistry, and the formation of aerosols, which affect the evolution, composition, and remote sensing of the planets (satellites). The upper atmosphere connects the planet (satellite) bulk composition to the near-planet (-satellite) environment. Understanding the relevant physics and chemistry provides insight to the past and future conditions of these objects, which is critical for understanding their evolution. This chapter introduces the basic concepts of upper atmospheres and ionospheres in our solar system and discusses aspects of their neutral and ion composition, wind dynamics, and energy budget. This knowledge is key to putting in context the observations of upper atmospheres and haze on exoplanets and to devise a theory that explains exoplanet demographics.

A. García Muñoz (✉)
Technische Universität Berlin, Berlin, Germany
e-mail: garciamunoz@astro.physik.tu-berlin.de

T.T. Koskinen
Lunar and Planetary Laboratory, University of Arizona, Tucson, AZ, USA
e-mail: tommi@lpl.arizona.edu

P. Lavvas
GSMA, UMR 7331, CNRS, Université de Reims, Champagne-Ardenne, Reims, France
e-mail: panayotis.lavvas@univ-reims.fr

,

## Contents

## Introduction

The fundamentals of upper atmospheres and ionospheres have been established in numerous works, including Bauer (1973), Rees (1989), and Schunk and Nagy (2000). Up-to-date views are presented in, e.g., Mendillo et al. (2002) and Nagy et al. (2008). The material covered in this chapter draws from these references as well as from others more specific. The scope of the chapter is that of aeronomy, which refers to the investigation of upper atmospheres and ionospheres as a subfield within the atmospheric sciences. For convenience, the text often refers to planets although it would be appropriate to refer to planets and their satellites.

Planetary atmospheres are vertically stratified, and it is common to differentiate various regions according to temperature or composition. Based on thermal structure, here the term upper atmosphere is used to encompass the thermosphere and exosphere. The thermosphere is typically heated by EUV and X-ray solar photons and shows a positive gradient of temperature with altitude up to an asymptotic value known as the exospheric temperature, $T_\infty$. Atop the thermosphere, in the exosphere gas particle collisions become rare, and particles with velocities larger than the gravitational escape velocity $v_\infty$ may leave to space. The exobase is the exospheric lower boundary and, by convention, occurs where $H/\lambda{\sim}1$, with $H$ being the atmospheric pressure scale height and $\lambda$ the gas mean free path.

Gas particles leave the exosphere in various ways. Thermal Jeans escape involves the dimensionless parameter $X_{exo}=GMm/R_{exo}kT_\infty$, formed as the ratio of the atom (or molecule) gravitational and thermal energies. $G$ is the gravitational constant, $M$ the planet mass, $m$ the particle mass, and $R_{exo}$ the distance from the planet center to the exobase. Jeans escape occurs for large $X_{exo}$ values. In this regime, the velocity distribution of gas particles remains essentially Maxwellian. Massive hydrodynamic escape occurs for $X_{exo}{\sim}2$–3 (Volkov et al. 2011), a condition easier to attain for light gases (H, $H_2$, He) at high temperatures and on low-mass planets.

Nonthermal escape processes include charge exchange, loss through open magnetic field lines, photoionization and dissociative recombination, and solar-wind pickup and sputtering. They may drive the escape of both neutrals and ions. The existence of a planetary magnetic field affects some of these processes, whose relative significance may change over the atmosphere's lifetime. Understanding the current escape processes is key to inferring a planet's history and predicting its evolution.

Based on neutral gas composition, the heterosphere refers to the altitudes for which molecular diffusion is more efficient in the transport of gases than eddy diffusion by large-scale winds and turbulent motions. Typically, the upper atmosphere overlaps with the heterosphere. The base of the heterosphere is at the homopause. Above it, the density of each long-lived gas drops according to its own scale height, which is inversely proportional to their corresponding mass. This separation by mass means that only the lighter gases reach the exosphere from where they can escape. Hydrodynamic escape conditions tend to facilitate the access of the heavier gases to the exosphere, thereby attenuating their separation by mass in the heterosphere.

The upper atmosphere is exposed to EUV (10–121 nm) and X-ray (0.01–10 nm) solar photons, as well as to cosmic rays and particles of auroral or solar-wind origin, all of which ionize the neutral gas and produce a weakly ionized plasma. Neutrals, ions, and electrons of planet origin coexist in the ionosphere and interact to some extent with the incoming solar wind. On planets without an intrinsic magnetic field, the ionopause sets the boundary between the dayside ionosphere and the solar wind and is perceived as an abrupt drop in the planet plasma density. The ionopause occurs as a balance between the solar-wind dynamic pressure and the planet plasma pressure and acts as an obstacle deflecting the incident solar wind. In the absence of an intrinsic magnetic field, the nightside ionosphere may extend as a tail on the planet shadow. For planets having a magnetic field, the ionosphere is contained inside the magnetosphere, and the planet plasma is confined by the field lines. Ions and electrons escape through open magnetic field lines, a process that is known as polar wind. Meteoritic material ablated as it enters the atmosphere may produce sporadic ionization layers.

Airglow and aurora are photoemission phenomena that offer unique opportunities for the remote sensing of upper atmospheres. They result from excited atoms and molecules radiating away their excess energy, thereby providing insight to the emitting gas (identity, abundance, production rate) and into the background atmosphere (density, temperature, velocity, energetic particle fluxes). The aurora is excited by precipitating electrons and ions from outside the atmosphere. Airglow is divided into day and night airglow (dayglow and nightglow, respectively). Sunlight is the ultimate excitation mechanism for airglow emissions, although the connection with solar photons is more direct for the dayglow than for the nightglow.

What follows reviews the aeronomy of the terrestrial planets (Earth, Venus, Mars), the gas giants (Jupiter, Saturn, Uranus, Neptune), and Saturn's moon Titan. The topic of ion exospheres is briefly mentioned in its application to Mercury and the Moon. We specifically acknowledge the authors of many seminal papers that have contributed greatly to the present knowledge of solar system aeronomy but that could not be cited here due to space limitations.

**Table 1** Summary of properties of the upper atmospheres and ionospheres of solar system planets and Titan

| | Homopause [altitude, km/pressure, bar] | Intrinsic magnetic field | Exobase [altitude, km] | $T_\infty$ [K] | $v_\infty$ [km/s][a] | Thermosphere: main gases | | Aurora | Some airglow emissions | |
|---|---|---|---|---|---|---|---|---|---|---|
| | | | | | | Neutrals | Ions | | Night | Day |
| **Earth** | ∼100/3 × $10^{-7}$ | Yes | ∼500 | 500–1000 | 11.2 | $N_2$, $O_2$, O, N, He | $NO^+$, $O_2^+$, $O^+$ | γ rays to radio frequencies | O($^1S$); 557 nm | $N_2(a)$; 140–180 nm |
| | | | | | | | | $O^+$; < 90 nm | $O_2(A,A',c)$; UV-NIR | O($^1D$); 630 nm |
| | | | | | | | | H, O, N, $O^+$, $N^+$, $H_2$, $N_2$; 90–200 nm | $O_2(a)$; 1270 nm | O($^1S$); 557 nm |
| | | | | | | | | | OH(X); Vis-IR | $O_2(b)$; 762 nm |
| | | | | | | | | | | $O_2(a)$; 1270 nm |
| **Venus** | ∼135/7 × $10^{-6}$ | No – ionopause at 225–375 km | ∼220–350 | ∼300 (day) /100 (night) | 10.4 | $CO_2$, $N_2$, O, CO | $O_2^+$, $CO_2^+$, $O^+$, $NO^+$ | | O($^1S$); 557 nm | CO(a); 190–260 nm |
| | | | | | | | | Diffuse, O($^3S^0$) and O($^5S^0$); 130.4 and 135.6 nm | $O_2(c)$; 400–700 nm | $CO_2^+(B)$; 289 nm |
| | | | | | | | | | $O_2(a)$; 1270 nm | O($^1S$); 557 nm |
| | | | | | | | | | OH(X); Vis-IR | $O_2(a)$; 1270 nm |
| | | | | | | | | | NO(C, A); 180–300, 1220 nm | |
| **Mars** | $\sim 130/3 \times 10^{-10}$ | Crustal | ~180–210 | ~200–350 (low/high solar activity) | 5.0 | $CO_2$, $N_2$, O, CO | $O_2^+$, $CO_2^+$, $O^+$, $NO^+$ | | NO(*C*, *A*); 180–300 | CO(*a*); 190–260 nm |
| | | | | | | | | Discrete CO(*A*), CO(*a*), $CO^+(B)$, $CO_2^+(B)$; 130–300 nm | OH (*X*); 1500–3000 nm | $CO_2^+(B)$; 289 nm |
| | | | | | | | | Diffuse: CO(*a*), $CO_2^+(B)$, $O(^1S)$ | $O_2(a)$ 1270 nm | $O(^1S)$; 297 nm |
| | | | | | | | | | | $O_2(a)$; 1270 nm |
| **Mercury** | | Yes | | | 4.3 | | $Na^+$, $O^+$ | | | Na; 589 nm |
| **Jupiter** | $\sim 10^{-6}$ | Yes | ~1600 | 900–1000 | 59.5 | $H_2$, He | $H^+$, $H_3^+$, $He^+$ | Main: H Lyman-α, $H_2$ UV bands, $H_3^+$ IR bands | H Lyman α | H Lyman-α, $H_2$UV bands (He 58.4 nm) |
| **Saturn** | $\sim 10^{-8} - 10^{-7}$ | Yes | ~2850 | 400–600 | 35.5 | $H_2$, He | $H^+$, $H_3^+$, $He^+$ | Main: | H Lyman-α | H Lyman-α |
| | | | | | | | | H Lyman-α | (He 58.4 nm) | $H_2$UV bands He 58.4 nm |
| | | | | | | | | $H_2$ UV bands | | |
| | | | | | | | | $H_3^+$ IR bands | | |

(*continued*)

**Table 1** (continued)

| | Homopause [altitude, km/pressure, bar] | Intrinsic magnetic field | Exobase [altitude, km] | $T_\infty$ [K] | $v_\infty$ [km/s][a] | Thermosphere: main gases | | Aurora | Some airglow emissions | |
|---|---|---|---|---|---|---|---|---|---|---|
| | | | | | | Neutrals | Ions | | Night | Day |
| **Uranus** | $\sim 10^{-4}$ | Yes | ~4700 | 850 | 21.3 | $H_2$, He | $H^+$, $H_3^+$, $He^+$ | H Lyman-α (?)<br>$H_2$ UV bands<br>$H_3^+$ IR bands | H Lyman-α | H Lyman-α<br>$H_2$UV bands |
| **Neptune** | $\sim 10^{-6}$ | Yes | ~2200 | 600 | 23.5 | $H_2$, He | $H^+$, $H_3^+$, $He^+$ | $H_2$ UV bands (?) | H Lyman-α | H Lyman-α, $H_2$UV bands |
| **Titan** | ~850 | No | ~1500 | ~150 | 2.6 | $N_2$, $CH_4$, $H_2$ | Main: $HCNH^+$, $C_2H_5^+$ | From Saturn's magnetosphere, similar emissions as from solar photons | From Saturn's magnetosphere, similar emissions as from solar photons | $N_2$: CY(0,0), LBH, VK NI: 120, 124.3, 149.3 nm NII: 108.5 nm |
| | | | | | | | + 100 s of other C/N/H/O composition ions | | | |

Additional refs.: [a]https://nssdc.gsfc.nasa.gov/planetary/factsheet/index.html

## Earth

The Earth has been investigated much more thoroughly than the other planets. Only a brief overview of its aeronomy is given here, to serve as background for Venus and Mars. The bibliography quoted above provides valuable references for more extensive treatments.

The Earth thermosphere extends from ~80 to ~500 km. Solar activity causes significant variability (on the order of hundreds of km) in the exobase level: higher activity implies additional energy deposited into the atmosphere, which expands as a consequence, and vice versa (Fig. 1). Smaller variations in the exobase level occur on diurnal timescales. For low (high) solar activity, usual exospheric temperatures are ~500 (~1000) K. The low abundance of $CO_2$ or other efficient IR radiators (e.g., NO) above the homopause (~100 km or $\sim 3 \times 10^{-7}$ bar) minimizes the thermostat effect that occurs on Venus and Mars and that prevents extreme exospheric temperature variations on these planets. Photodissociation of $N_2$ and $O_2$ (main neutrals in the bulk atmosphere) and molecular diffusion result in O and N

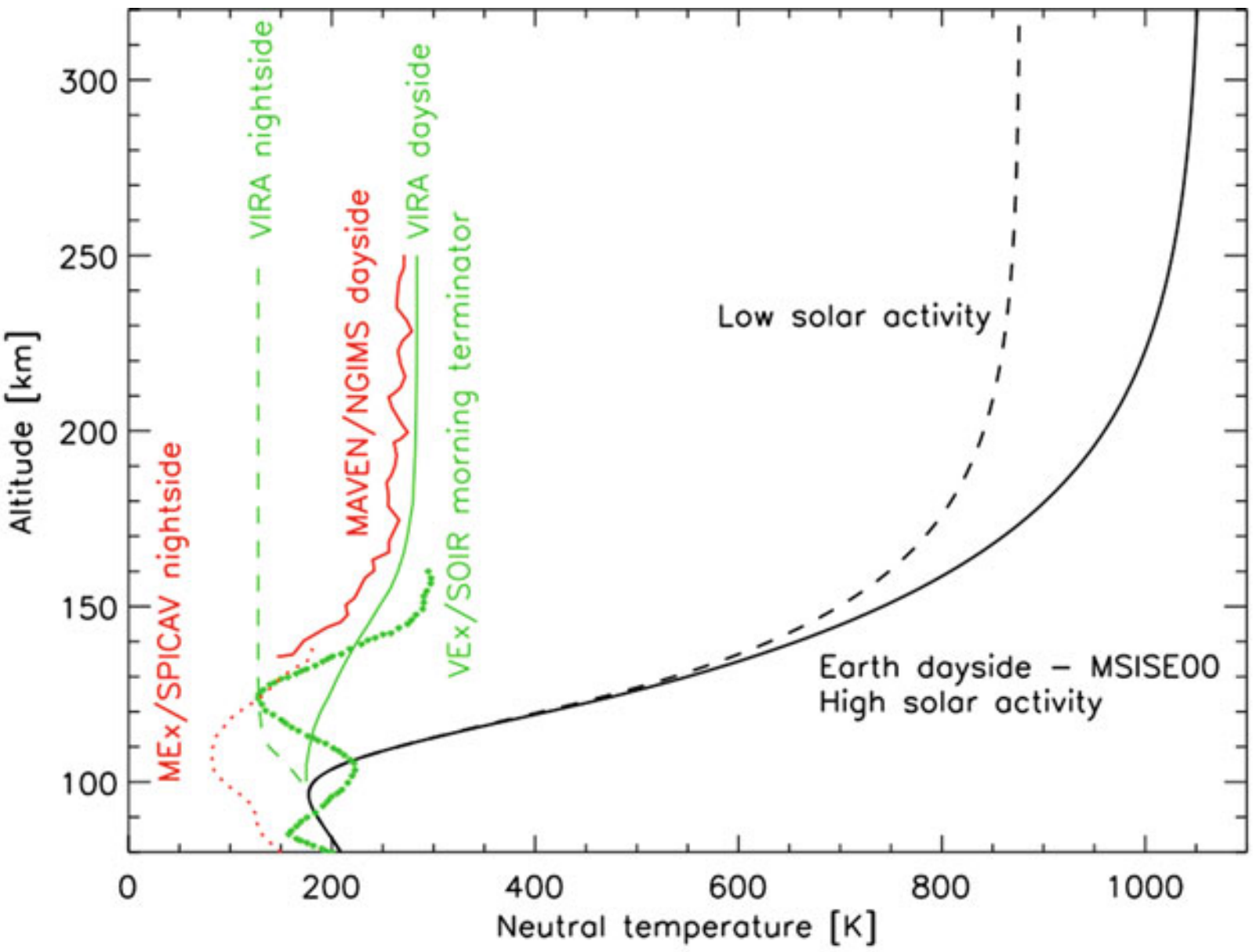

**Fig. 1** Thermospheric temperatures for the Earth, Venus, and Mars. *Black*: Earth profiles for dayside conditions, as obtained from the NRLMSISE-00 model (http://ccmc.gsfc.nasa.gov/modelweb/models/nrlmsise00.php). *Green*: Venus profiles for day- and nightside conditions (VIRA model, moderate solar activity; Keating et al. 1985) and for morning terminator conditions (solar occultation measurements; Mahieux et al. 2015). *Red*: Mars profiles for dayside (Bougher et al. 2015b) and nightside conditions (stellar occultation measurements; Forget et al. 2009). Error bars are omitted

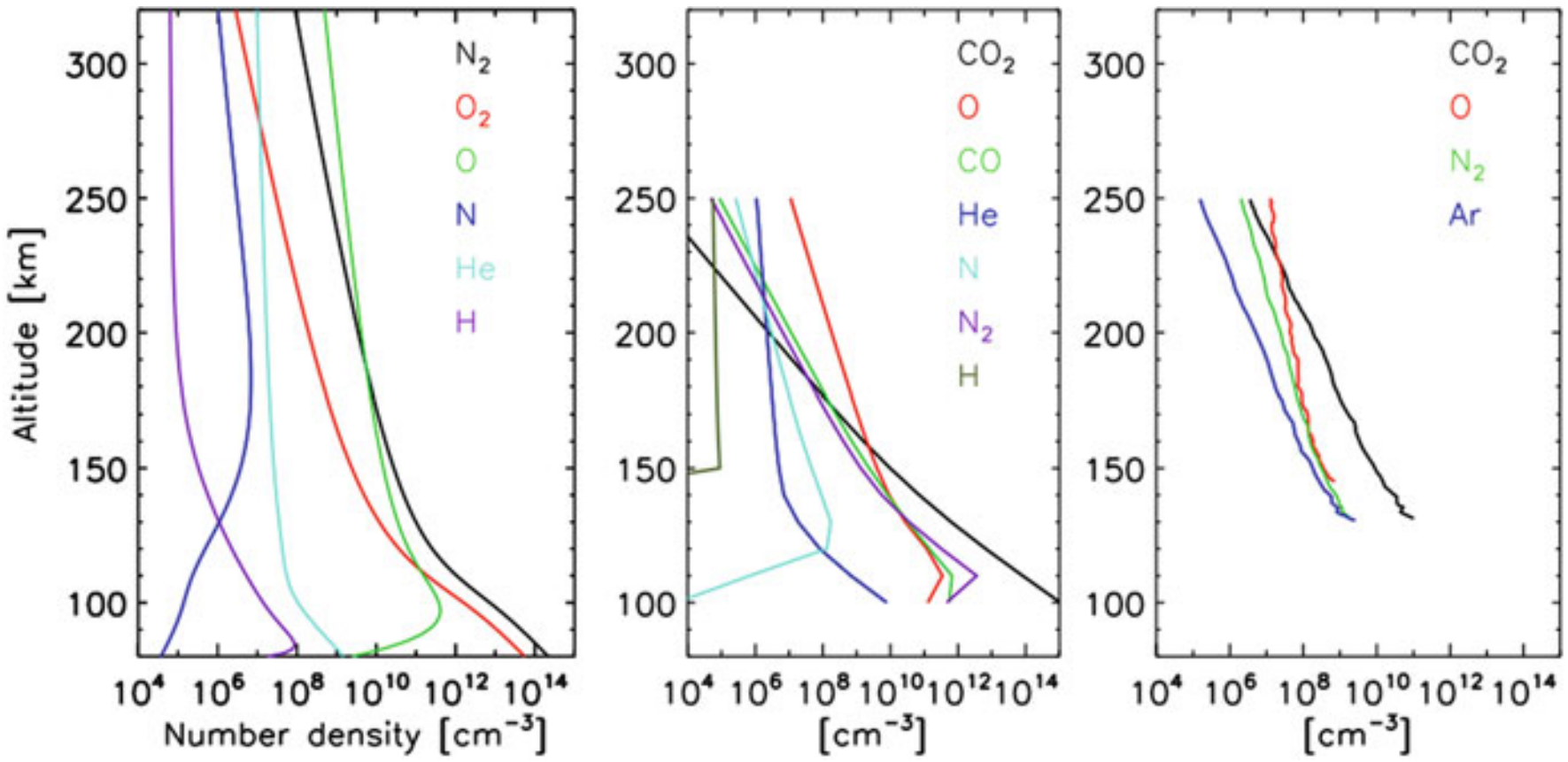

**Fig. 2** Neutral gas composition of the Earth (*left*, NRLMSISE-00 model at high solar activity, dayside), Venus (VIRA model at moderate solar activity; Keating et al. 1985), and Mars (MAVEN/NGIMS measurements; Bougher et al. 2015b). Error bars are omitted

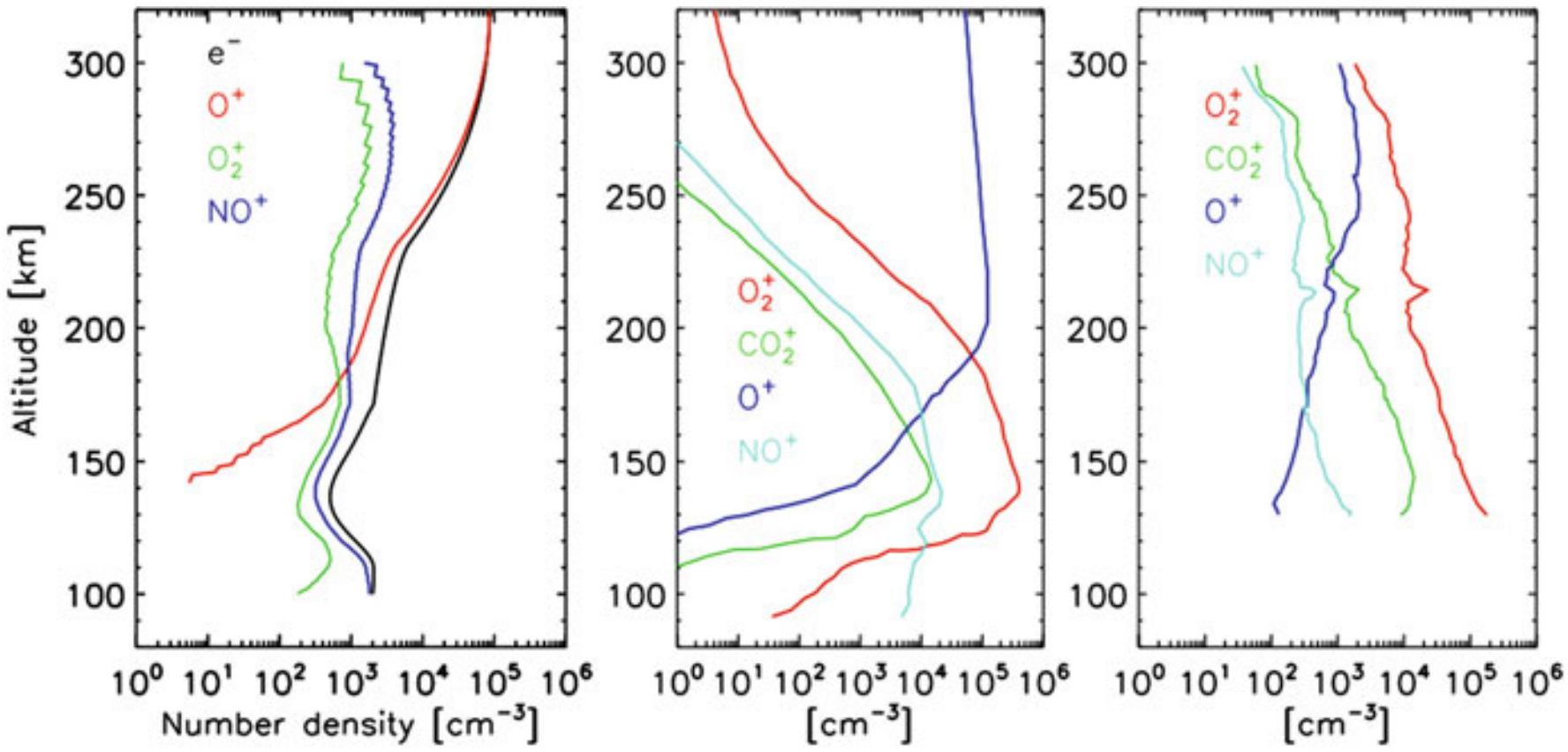

**Fig. 3** Dayside ionosphere of the Earth (*left*, IRI-2007 model; http://ccmc.gsfc.nasa.gov/modelweb/models/iri_vitmo.php), Venus (photochemical model; Fox and Sung 2001), and Mars (Bougher et al. 2015b). Error bars are omitted

atoms (with He) becoming abundant in the thermosphere (Fig. 2). In the exosphere, the lighter gases H and He prevail.

Earth's ionosphere is traditionally split into D, E, $F_1$, and $F_2$ layers, ordered from lower to higher altitude. This denomination has guided the naming of other ionospheres. Molecular ions ($NO^+$, ${O_2}^+$) dominate in the D and E layers, whereas $O^+$, the product of O photoionization, tends to dominate in the F layers (Fig. 3). Peak electron densities occur in the $F_2$ layer at and above which ion diffusion becomes important.

Hydrodynamic escape is currently ineffectual on Earth but likely to have occurred in the past, especially if exospheric hydrogen was abundant and the young Sun's EUV output stronger (Hunten 1993; Catling and Zahnle 2009). The impact of nonthermal escape over the planet lifetime is sensitive to the existence of a magnetic field in the early Earth, a possibility that adds uncertainty to the evolution of the terrestrial atmosphere (Lammer et al. 2008).

Earth shows a variety of airglow and auroral emissions sharing both commonalities and differences with the Venus and Mars emissions (Meier 1991; Galand and Chakrabarti; Slanger and Wolven, both in Mendillo et al. 2002). Some similarities arise from the fact that O and N atoms are produced on all three planets by photodissociation of $O_2/N_2$ (Earth) and $CO_2/N_2$ (Venus, Mars). The lack of an intrinsic magnetic field on Venus and Mars causes discrepancies in their aurora excitation with respect to Earth.

## Venus

The Venus homopause lies at ~135 km altitude (~$7 \times 10^{-6}$ bar), the exact level being lower for light gases (H, $H_2$, He) and higher for the heavy ones. The thermosphere extends from 120 to 220–350 km, and its neutral composition is dominated by $CO_2$ and $N_2$ up to 140–160 km (Fig. 2). Above, $CO_2$ photodissociation and diffusion cause O and CO to become dominant. In the exosphere, H and He are major constituents.

Venus' ionosphere is also structured in layers that reveal changes in electron densities and ionizing processes (Pätzold et al. 2007). The secondary ionization layer rests at the base of the ionosphere at ~120 km. Above it, the main ionization layer reaches electron densities of a few times $10^5$ $cm^{-3}$ at its ~140 km peak. These layers are caused by soft X-ray (secondary) and EUV (main) photon photoionization, respectively, and tend to form $CO_2^+$. Rapid reaction of $CO_2^+$ with O leads to $O_2^+$ as the main ion up to ~180 km (Fig. 3). The $O_2^+$ ion is lost in the dissociative reaction with electrons, forming two O atoms. A third ionization layer dominated by $O^+$ occurs above ~180 km. A sporadic ionization layer attributed to ablating meteoroids occurs near 115 km, which may consist of $Mg^+$ and $Fe^+$ ions (Withers 2012).

Electron densities on Venus are highly variable in the topside ionosphere and decay abruptly near the ionopause. For solar minimum conditions, the ionopause level fluctuates between 225 and 375 km within a few days. This variability reflects the ever-changing interaction between the solar-wind and planet plasma. On the nightside, a weak ionosphere occurs sustained by $O^+$ transported from the dayside and precipitating electrons. At times, the nightside ionosphere nearly vanishes (Cravens et al. 1982).

Fast thermospheric winds participate in a subsolar-to-antisolar (SS-AS) circulation above ~100 km driven by dayside solar heating that produces a day-to-night pressure gradient (Fox and Bougher 1991; Clancy et al. 2015). Upwelling and downwelling occur near the subsolar and antisolar points, respectively. Below $\sim$100 to 120 km, the SS-AS circulation connects with the super-rotating flow that dominates Venus' lower-atmosphere circulation. It takes a few days for the O and N atoms formed by $CO_2$ and $N_2$ photodissociation to be transported to the nightside and recombine, resulting in $O_2$, NO, and OH nightglow at UV-to-NIR wavelengths and 90–130 km altitudes (García Muñoz et al. 2009). These emissions are variable and asymmetric with respect to the antisolar point, which suggests complex wind interactions and competing quenching-vs-radiation effects.

Venus' dayglow includes emissions from He, $He^+$, H, O, $O^+$, N, C, $C^+$, CO, and ${CO_2}^+$, whose interpretation requires elaborate modeling of different excitation processes (Fox and Sung 2001). A diffuse oxygen aurora exists on the nightside, possibly excited by precipitating energetic electrons. It is unclear what drives the precipitation in the absence of an intrinsic magnetic field, although magnetic reconnection might provide such a mechanism (Zhang et al. 2012). The intermittent oxygen green line nightglow, potentially correlating with solar activity, may prove key to resolving some of these uncertainties (Slanger et al. 2001). X-ray emission, whether the result of fluorescent scattering of solar X-rays or charge exchange interactions with the solar wind, provides a complementary and as-of-yet little explored view of the Venus thermosphere-exosphere and their interaction with the Sun (Dennerl 2008).

The ion and electron temperatures reach thousands of degrees in the exosphere, thus exceeding the neutral temperature over most of the ionosphere (Miller et al. 1980). The neutral thermosphere is relatively cold with temperatures of $\sim$300 K at the dayside exobase, much less than at Earth (Fig. 1). On the nightside, thermospheric temperatures of $\sim$100 K are the lowest on Venus, and this region is often referred to as the cryosphere. Radiation at 15 $\mu$m from $CO_2$ (a trace gas in the terrestrial atmosphere but abundant on Venus and Mars) in nonlocal thermodynamical equilibrium efficiently cools the Venus thermosphere and attenuates the potentially larger impact of solar activity at an inner planet. The day-night thermal contrast is coupled with the SS-AS circulation. Waves associated with density modulations (Müller-Wodarg et al. 2016), possibly originating from the lower atmosphere, have been reported in the thermosphere.

Escape from the Venus upper atmosphere is presently dominated by nonthermal processes (Lammer et al. 2008; Catling and Zahnle 2009), although thermal (hydrodynamic) loss may have been significant in the past. Indeed, the ancient Venus may have experienced a runaway greenhouse effect that resulted in abundant upper-atmosphere steam. Exposed to EUV sunlight, the water would dissociate, with the H atoms thermally escaping more easily than the heavier O atoms. A mass-based fractionation would also occur for the D isotope, which might explain the D/H ratio of $(1.6–2.2) \times 10^{-2}$, two orders of magnitude larger than at Earth. Hydrodynamic escape may have removed a water ocean (if it existed) in less than 100,000 years. This example highlights the importance of the upper atmosphere for planetary evolution.

## Mars

Both Mars and Venus lack intrinsic magnetic fields, although Mars does have a leftover crustal field after its presumed original intrinsic field vanished 4 gigayears ago (Mangold et al. 2016). This remnant field is apparent over localized near-surface areas of Mars and affects aspects of its aeronomy such as the ionospheric structure, interaction with the solar wind, and aurora.

The Martian homopause lies at ~130 km (~$3 \times 10^{-10}$ bar). Lower down, $CO_2$ and $N_2$ are well mixed and dominate the background composition (Fig. 2). Above, CO and O are in diffusive equilibrium and become locally abundant. O takes over as the main atmospheric gas for altitudes larger than ~200 km, near the exobase (Bougher et al. 2015a, b; Bhardwaj et al. 2016).

The bottom boundary of the Mars dayside thermosphere lies at ~100 km, where the temperature is ~120 K (Fig. 1). The temperature rises to exospheric values of ~200 and 350 K for low and high solar activity conditions, respectively (Mueller-Wodarg et al. 2008; Bougher et al. 2015a). Thermal conduction and $CO_2$ cooling at 15 μm, excited in collisions with O, balance the heating by EUV solar energy deposition. The fact that the temperature fluctuations in the Martian thermosphere are larger than for Venus suggests that the $CO_2$ thermostat is less efficient on Mars, probably because the Martian $O/CO_2$ density ratio is smaller. The magnitude of the day-night temperature variations in the thermosphere can be up to ~200 K depending on solar activity. Thermospheric winds transport heat from the dayside to polar latitudes and on to the nightside (Bougher et al. 2015a; González-Galindo et al. 2015). They also transport the O and N products of $CO_2$ and $N_2$ photodissociation that recombine on the nightside to produce nightglow (e.g., Clancy et al. 2013).

The lower and upper atmospheres of Mars are strongly coupled. The thermosphere is forced by upward-propagating tides and gravity waves originating near the surface, possibly affected by topography. Dust storms cause transient heat deposition that modifies these wave interactions and result in thermospheric density changes by factors of a few (Withers and Pratt 2013).

Electron densities in the dayside ionosphere have been measured from ~80 to 500 km. Peak densities are a few times $10^5$ $cm^{-3}$ at 120–140 km within the main ionization layer, which is ionized by EUV solar photons (Fig. 3). A secondary layer is formed at ~100 km by soft X-ray solar photons. Similar to Venus, $O_2{}^+$ is the dominant ion in the main ionization layer, and $NO^+$ is predicted to contribute to the secondary ionization layer (Krasnopolsky 2002; Fox 2009). The $O_2{}^+$ and $O^+$ densities become comparable at 250–300 km. Peak electron densities on the nightside are typically two orders of magnitude smaller than on the dayside, consistent with an origin due to plasma transport from the dayside and electron precipitation (Withers et al. 2012). Abrupt drops in dayside electron density characteristic of ionopause-like configurations have been reported (Vogt et al. 2015) and seem to be sensitive to solar-wind conditions and to the local crustal magnetic field.

Mars dayglow includes emissions from $N_2$, CO, $CO_2^+$, C, O, $H_2$, and H that originate from altitudes up to ~200 km. The gases are excited either directly (e.g., $CO_2$+photon leading to excited states of O, $CO_2^+$, CO, $CO^+$) or indirectly via neutral and ion chemistry (e.g., $O_2^+ + e \rightarrow O(^1S) + O$). The scale heights for each emission profile reflect the details of the primary (and secondary) excitation mechanisms and are used to infer neutral densities and exospheric temperatures (Huestis et al. 2010).

Precipitating electrons of energies 300–1000 eV are accelerated by the crustal field and produced upon collision with the neutral atmosphere sporadic aurorae at ~130 km that resemble Earth's discrete aurora (Bertaux et al. 2005). Mars also exhibits a planetwide diffuse aurora that reaches down to 60 km and is likely excited by solar particles of hundreds of keV (Schneider et al. 2015).

The moderately elevated D/H ratio in the Martian atmosphere (a few times that of Earth) and the evidence for past liquid on its surface suggest that the planet may have lost a substantial amount of its initial water (Hunten 1993). The possibility for surface water shows the contrast between Mars' early (wet, warm) and current (dry, cold) climates. On current Mars, thermal (Jeans) escape is relevant for the loss of hydrogen, whereas nonthermal escape contributes to the loss of heavier atoms (Lammer et al. 2008; Catling and Zahnle 2009). Hydrodynamic escape is not operating now but is likely to have operated in the past. Bursts of solar activity, as during coronal mass ejections, were more frequent for the early Sun than they are now and may have significantly enhanced the past escape rates (Jakosky et al. 2015).

The recent detection of a transient extended brightness feature of uncertain physical origin at the Mars morning terminator and ~250 km altitude (Sánchez-Lavega et al. 2015) shows that there remain significant uncertainties in our understanding of the Mars upper atmosphere. Three space missions (ExoMars, MAVEN, MOM) have recently entered into Mars orbit and are contributing to a better understanding of Martian aeronomy. Some mission highlights include the discovery of metal ion layers of $Na^+$, $Mg^+$, and $Fe^+$ (Grebowsky et al. 2017) or a better constraint on the atmospheric Argon fractionation and, in turn, the prediction that about two-thirds of this noble gas (and a sizeable amount of the bulk $CO_2$) may have been lost to space over the planet history (Jakosky et al. 2017).

## Ion Exospheres

Ionospheres occur also on bodies with tenuous atmospheres such as Mercury or the Moon. These ion exospheres contain the signature of the surface and interior material that is being released.

Mercury's ionosphere contains $Na^+$ and $O^+$ concentrated preferentially near the magnetic poles, which points to these regions as sources of the heavy ions probably through solar-wind sputtering. The lighter ion $He^+$ is observed more uniformly around the planet, and its distribution is consistent with planetwide evaporation from the surface (Zurbuchen et al. 2011).

There is evidence that the near-Moon environment is partly ionized and that electron densities can reach values of $10^3$ $cm^{-3}$ (Choudhary et al. 2016). Modeling suggests that the measured plasma is consistent with molecular ions of $H_2O^+$, $CO_2{}^+$, and $H_3O^+$ rather than inert ions ($Ar^+$, $Ne^+$, $He^+$). Other interpretations suggest that the Moon's ion exosphere is caused by electron emission from dust (Stubbs et al. 2011).

## The Upper Atmospheres of Jupiter, Saturn, Uranus, and Neptune

Giant planet thermospheres are composed of $H_2$ and He with some H and traces of carbon and oxygen species (Fig. 4). Methane is the dominant carbon-bearing species, and its abundance falls off rapidly with altitude above the homopause. The abundance of He also decreases above the homopause. Atomic H is mostly released by photochemistry below the thermosphere, and, being lighter than $H_2$, its abundance increases with altitude in the thermosphere. An external flux of water group species has been inferred for all of the giant planets (Feuchtgruber et al. 1997), and on Saturn, water "raining" down from the magnetosphere and rings affects the ionosphere (e.g., Connerney and Waite 1984; O'Donoghue et al. 2013). The abundance of water is roughly constant in the thermosphere and decreases with pressure in the lower atmosphere due to condensation (Moses and Bass 2000; Müller-Wodarg et al. 2012). The dominant ions in the main ionospheric peak are $H^+$ and $H_3{}^+$.

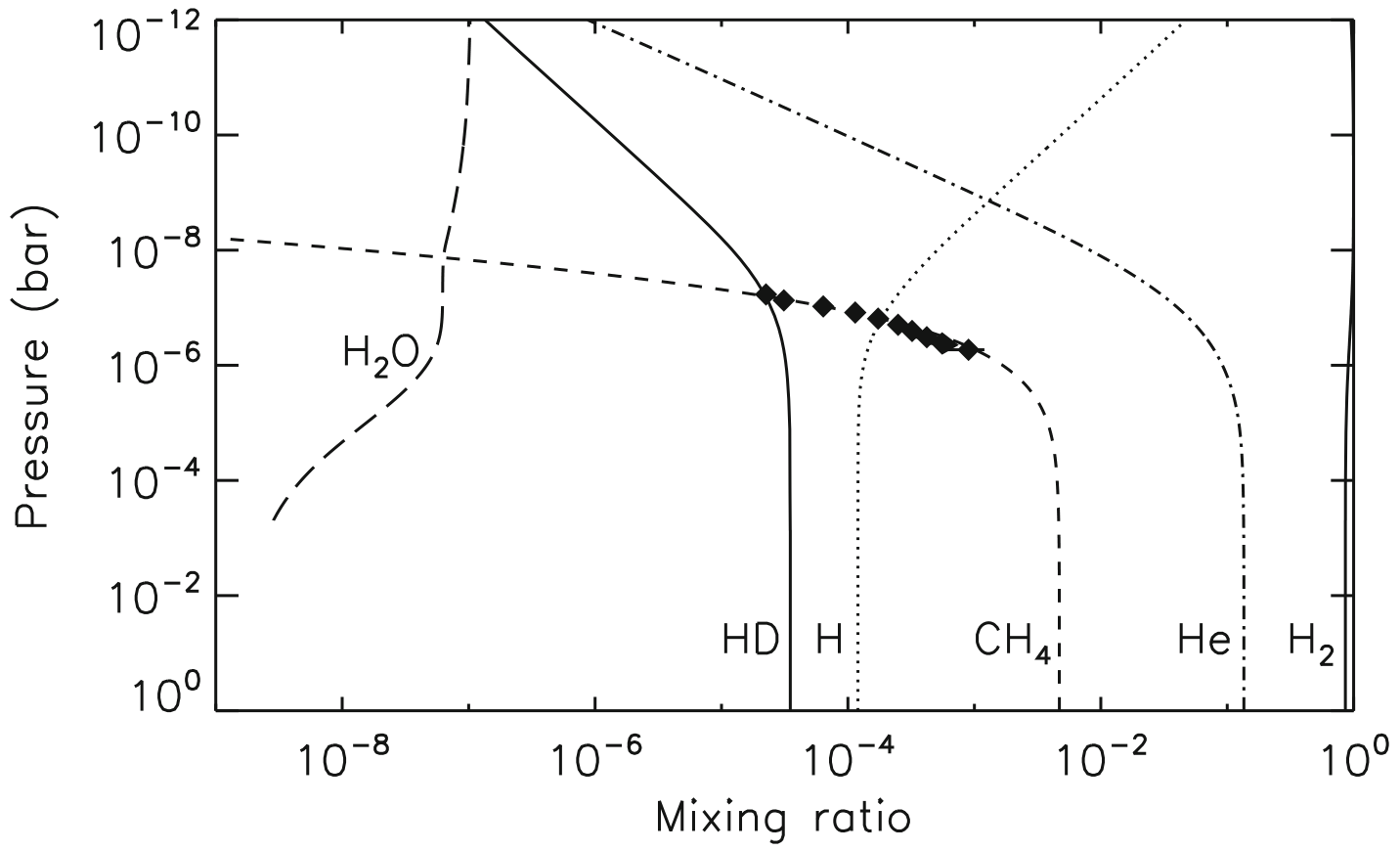

**Fig. 4** Mixing ratios in Saturn's atmosphere illustrate the basic composition of giant planet upper atmospheres (Strobel et al. 2016). The data points (*diamonds*) were retrieved from a Cassini/UVIS stellar occultation

## Observations

Our understanding of giant planet upper atmospheres is mostly based on the Pioneer, Voyager, Galileo, and Cassini space missions, although observations by Earth-orbiting space telescopes and ground-based telescopes in the near-IR have also contributed. The density and temperature profiles in Jupiter's equatorial thermosphere were measured by the Galileo probe (Seiff et al. 1998). In situ measurements of Saturn's thermosphere are also planned for the Cassini Grand Finale in 2017 (Edgington and Spilker 2016). All other information comes from remote sensing. In particular, UV solar and stellar occultations observed by visiting spacecraft are an important tool for retrieving densities of $H_2$ and hydrocarbons as well as temperatures in the upper atmosphere. Occultations by the Voyager Ultraviolet Spectrometer (UVS) have been analyzed on all giant planets (e.g., Yelle et al. 1993; Stevens et al. 1993; Yelle and Miller 2004; Vervack and Moses 2015), the Cassini Ultraviolet Imaging Spectrograph (UVIS) has spent a decade observing them on Saturn (Koskinen et al. 2013, 2015, 2016), and the New Horizons (NH) ALICE instrument also observed stellar occultations by Jupiter (Greathouse et al. 2010).

Ultraviolet aurora and airglow, including H Lyman-α (H Lyα), $H_2$ electronic band, and He 584 Å emissions, are also used to study the upper atmosphere. The aurora is excited by electron precipitation along magnetic field lines connecting the polar ionosphere to the solar-wind interaction region and the magnetosphere. Voyager/UVS obtained the first unambiguous detections of the UV aurora on Jupiter, Saturn, and Uranus and found evidence of the aurora on Neptune. Subsequent observations by the HST and Cassini have been used to study the morphology and physical origin of giant planet aurora (e.g., Grodent 2015). The aurora on Jupiter and Saturn has also been detected at visual wavelengths, and X-ray emissions from the aurora and disk have been detected on Jupiter (e.g., Badman et al. 2015). The primary origin of the H Lyα and He 584 Å airglow is resonant scattering of sunlight (e.g., Ben-Jaffel et al. 1995; Parkinson et al. 1998). The $H_2$ band emissions are probably produced by a combination of photoelectron excitation and solar fluorescence (e.g., Liu and Dalgarno 1996), although some authors argue that additional excitation by suprathermal electrons or "electroglow" is required to explain these emissions (e.g., Herbert and Sandel 1999).

Near-IR emissions from the upper atmosphere consist of 3.3 μm $CH_4$ emissions that originate near the homopause (Drossart et al. 1999) and emissions at 2–4 μm from the thermosphere (Drossart et al. 1989). Observations of $H_3{}^+$ emissions probe the aurora, the state of the ionosphere, temperature, and dynamics (Miller et al. 2006, 2010). Emissions from the aurora and disk are observed on Jupiter and Uranus, while on Saturn, auroral emissions are observed regularly, and disk emissions appear intermittently (O'Donoghue et al. 2013). Unfortunately, there are at present no means to detect any other ions. The only other information on the ionosphere are electron densities retrieved from spacecraft radio occultations. They have been retrieved for Jupiter and Saturn from Pioneer data, for all giant planets from Voyager data (Kliore et al. 1980; Lindal et al. 1985, 1987; Lindal 1992; Yelle and Miller 2004), for Jupiter from Galileo data (Hinson et al. 1997), and for Saturn from Cassini data (Kliore et al. 2009).

## Thermospheres

The temperature in the stratosphere and mesosphere is controlled by solar near-IR heating in $CH_4$ bands and IR emissions by $CH_4$ and photochemical products $C_2H_6$ and $C_2H_2$ (Yelle et al. 2001). As the abundance of $CH_4$ decreases above the homopause (Fig. 4), the lack of radiative cooling allows for a hot thermosphere. Unlike on Earth, on the giant planets, the thermospheres are much hotter than expected from solar heating (Fig. 5), and the solution to this "energy crisis" remains elusive (see below). The upper atmosphere of Neptune is slightly warmer than on Saturn, although generally the temperatures on these two planets appear comparable. The temperatures on Jupiter and Uranus, on the other hand, are much higher than on Saturn and Neptune. These trends do not correlate with distance from the Sun, and, in the absence of a definite solution to the energy crisis, there is no generally accepted explanation for these differences.

The location of the base of the thermosphere should coincide roughly with the homopause, i.e., the region where the abundance of $CH_4$ begins to fall rapidly with altitude. On Jupiter, the stratospheric mixing ratio of methane is $1.8 \times 10^{-3}$, and the homopause is near the 1 μbar level, close to the base of the thermosphere (Seiff et al. 1998). The stratospheric $CH_4$ mixing ratio of $4.8 \times 10^{-3}$ on Saturn (Fletcher et al. 2009) is the highest among the giant planets. At low latitudes, the homopause on Saturn is near 0.01–0.1 μbar, in rough agreement with the base of the thermosphere (Koskinen et al. 2015). In contrast to Jupiter and Saturn, temperatures on Uranus and Neptune are cold enough for methane to condense in the troposphere. On Uranus, the stratospheric mixing ratio of $CH_4$ is only $10^{-5}$–$10^{-4}$, and the abundance decreases further above the 0.1 mbar level (e.g., Lellouch et al. 2015), possibly explaining the relatively deep base of the thermosphere on Uranus. On Neptune, the stratospheric $CH_4$ mixing ratio of about $10^{-3}$ is several times higher than allowed by the mean tropospheric cold trap. This is either because of a leakage through a warm tropopause at high southern latitudes or upwelling and convective overshooting (e.g., Lellouch et al. 2015). The homopause is near the 1 μbar level, roughly in line with the base of the thermosphere (Yelle et al. 1993; Moses et al. 1995).

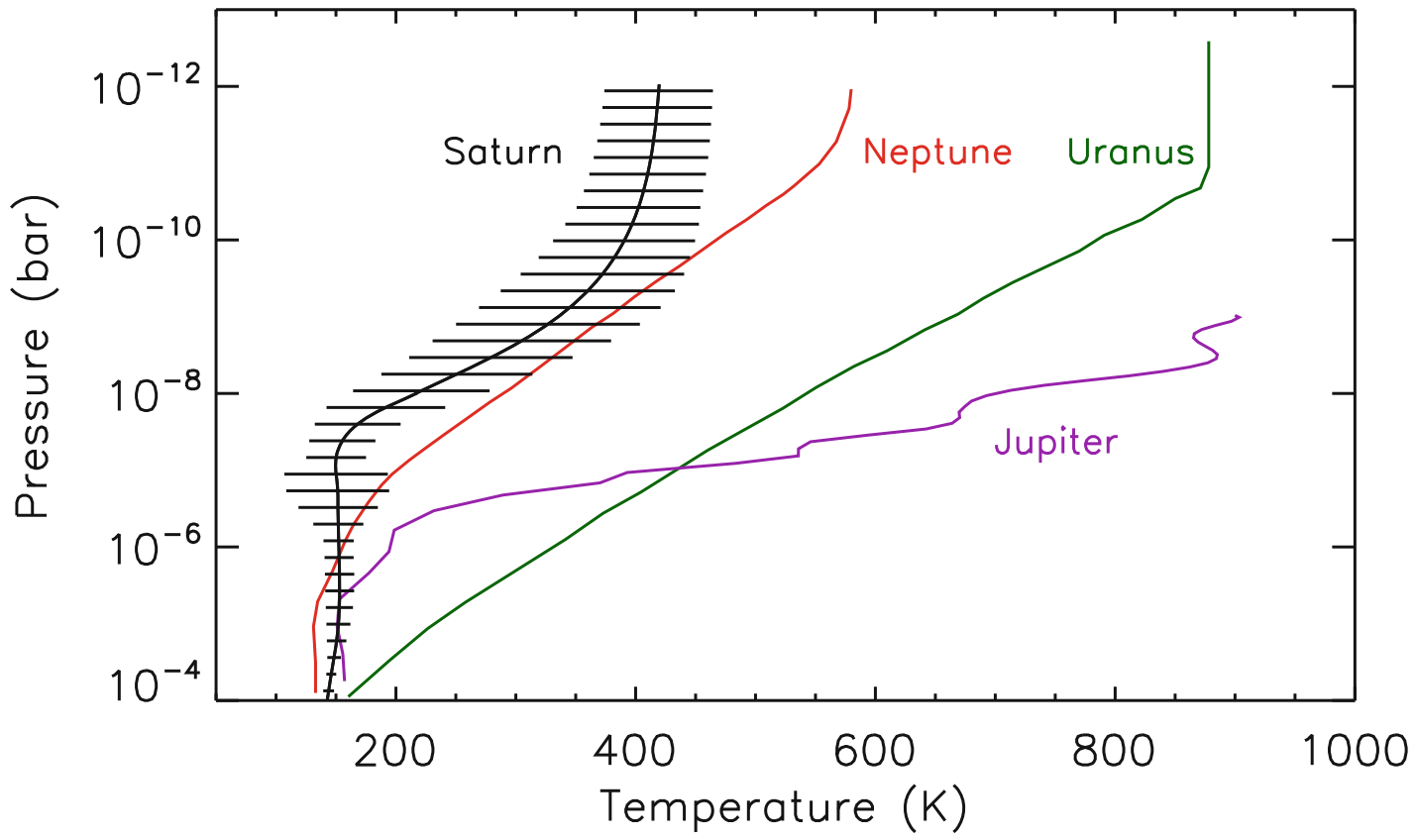

**Fig. 5** Low-latitude temperature-pressure (T-P) profiles for Jupiter from the Galileo probe (Seiff et al. 1998) and Uranus from the Voyager 2/UVS solar occultation (Stevens et al. 1993). The Saturn T-P profile is an average of 28 low to mid-latitude Cassini stellar occultations combined with Cassini/CIRS data (Koskinen et al. 2015), with error bars reflecting the variability of the observations. The T-P profile for Neptune is based on the Voyager 2/UVS occultations (Müller-Wodarg et al. 2008). The highest temperature on Jupiter expected from solar XUV heating is only 230 K

Most of the work on the energy crisis has focused on Jupiter and Saturn where more observations are available. The commonly proposed solutions are the deposition of energy by breaking gravity and acoustic waves or resistive (Joule) heating by auroral electrodynamics followed by redistribution of energy by global circulation (e.g., Müller-Wodarg et al. 2006). There are, however, problems associated with both of these solutions. While wave heating has been proposed as a plausible mechanism on Jupiter (e.g., Young et al. 1997; Schubert et al. 2003), the calculations to date are idealized and ignore, for example, momentum deposition by waves, which would considerably decrease their energy flux (Yelle and Miller 2004). Similarly, wave heating has been found to be insufficient on Saturn. In order to be significant, wave heating would also have to be globally distributed and continuously active, which may be unlikely (Strobel et al. 2016).

Magnetosphere-ionosphere coupling generates auroral electric currents that power resistive heating of the polar upper atmosphere and ionosphere. The derived resistive heating rates of about 100 TW on Jupiter and about 10 TW on Saturn are in principle sufficient to solve the energy crisis, but the heating is limited to a narrow band of latitudes near the poles (e.g., Müller-Wodarg et al. 2006). Majeed et al. (2005) used a circulation model to argue that meridional winds could transport energy to low latitudes on Jupiter and explain the equatorial temperatures. More recent modeling on Jupiter and Saturn, however, demonstrates that westward ion drag and a "Coriolis barrier" imposed by rapid rotation turn meridional winds into a strong high latitude zonal jet, thus preventing the redistribution of energy to the equator (Smith et al. 2007; Smith and Aylward 2009). In contrast, new data from Cassini show that on Saturn the poles are generally warmer than the equator (Koskinen et al. 2015), indicating that polar heating and redistribution to lower latitudes may in fact be operating. No definite mechanism to facilitate the redistribution of energy, however, has yet been identified and the possibility that some other heating mechanism operates at low to mid-latitudes cannot be ruled out.

## Ionospheres

In principle, the ionospheres of the giant planets should be simple because the atmospheres are dominated by $H_2$ and He. According to the basic theory, solar UV radiation and electron precipitation ionize $H_2$, producing $H_2^+$ that rapidly reacts with $H_2$ to form short-lived $H_3^+$, which recombines dissociatively with electrons to release $H_2$ and H. Ionization of H and dissociative ionization of $H_2$ form the long-lived $H^+$, while ionization of He produces $He^+$, which can also react with $H_2$ to produce small amounts of $HeH^+$ (e.g., Yelle and Miller 2004). Ionization of $CH_4$ near the homopause leads to the production of complex, short-lived hydrocarbon ions and heavier neutral molecules, including $C_6H_6$, that can act as a stepping stone to ring polyaromatic hydrocarbons and eventually stratospheric haze (Kim and Fox 1994; Friedson et al. 2002; Wong et al. 2003; Kim et al. 2014; Koskinen et al. 2016).

This basic theory is undoubtedly correct, and yet models have struggled to match the observed electron densities. Figure 6 compares electron density profiles retrieved from radio occultations. The results indicate strong variability in electron densities on Jupiter and Saturn that is not clearly understood (e.g., Yelle and Miller 2004; Kliore et al. 2009). Similar variability may occur on Uranus and Neptune, but observations are more limited. The observed profiles also include sharp, dense layers that can be driven by waves (Matcheva et al. 2001). Assuming that photoionization dominates at non-auroral latitudes, the electron densities should decrease with distance from the Sun. Figure 6 confirms that the overall electron density decreases

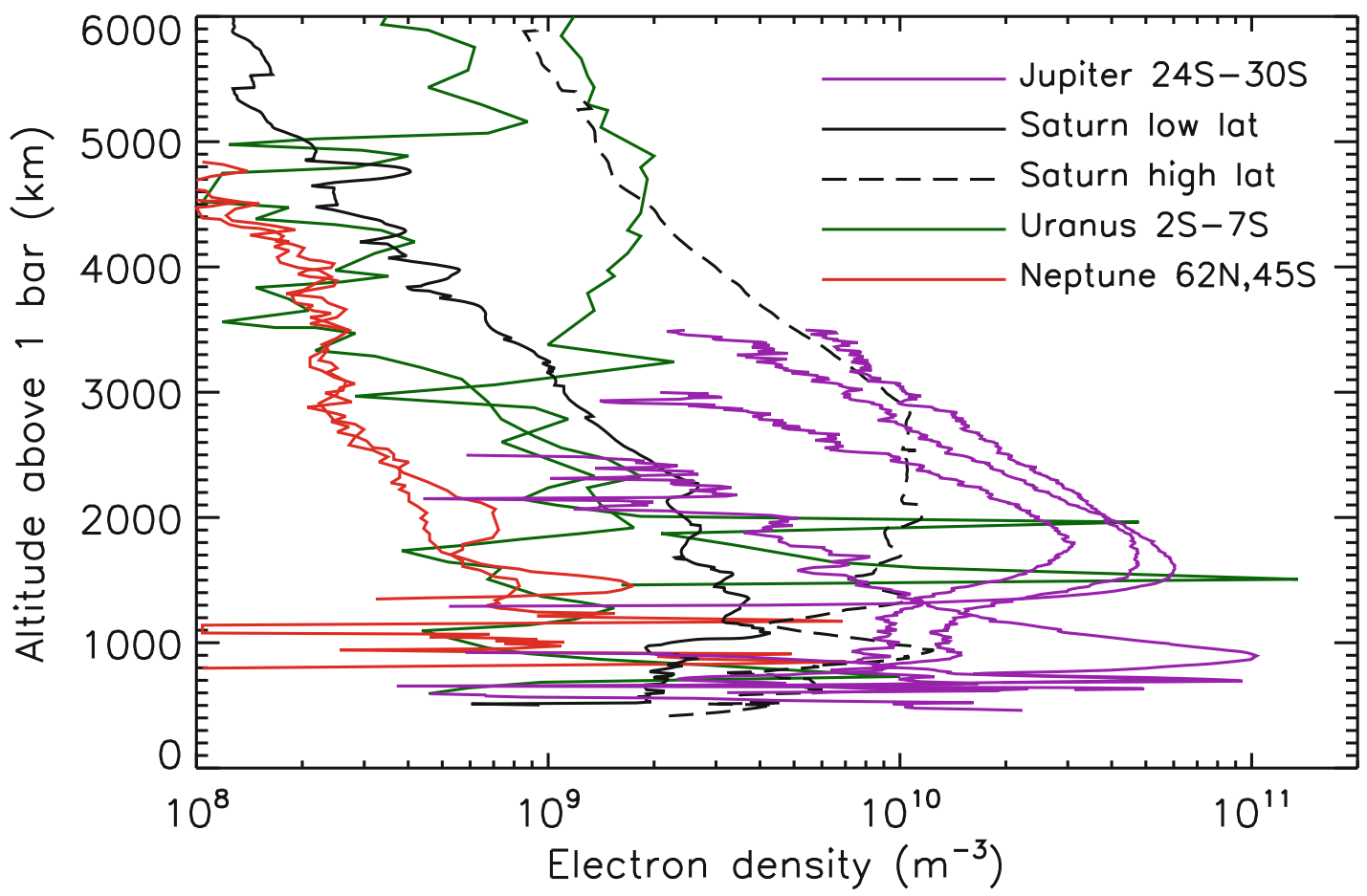

**Fig. 6** Electron density profiles in giant planet ionospheres retrieved from radio occultations. The results for Jupiter are from Galileo (Hinson et al. 1997), available through the Planetary Data System. The Saturn low-latitude results are an average of 17 occultations within 30° latitude from the equator, and the high latitude results are an average of 12 occultations at absolute latitudes higher than 40° (Kliore et al. 2009). The Voyager ingress and egress results for Uranus and Neptune were taken from Lindal et al. (1987) and Lindal (1992), respectively

from Jupiter through Saturn to Neptune. The situation on Uranus appears more complicated, given the sharp high-density peaks in the lower ionosphere and high altitude electron densities that exceed equatorial densities on Saturn.

According to models, which focus mostly on Jupiter and Saturn, the lower ionosphere above the hydrocarbon layer is dominated by $H_3{}^+$, while $H^+$ dominates at higher altitudes. Simple models predict that the transition from $H_3{}^+$ to $H^+$ takes place below the main ionospheric peak and underestimate the altitude of the peak while significantly overestimating the electron densities. Several possible solutions to these problems have been proposed. On Jupiter, models that include vertical plasma drifts and vibrational excitation of $H_2$ have been used to match the observed electron densities (e.g., Majeed et al. 1999; Yelle and Miller 2004). Plasma drifts move the ionospheric peak to higher altitudes, while reactions of vibrationally excited $H_2$ effectively convert $H^+$ to $H_3{}^+$, allowing for lower electron densities. On Saturn, models that invoke an influx of water from the magnetosphere and rings have been used to achieve conversion of $H^+$ to $H_3{}^+$ and an agreement with the observed electron densities (e.g., Müller-Wodarg et al. 2012). In both cases, solar photoionization is sufficient to explain the non-auroral electron densities.

Impact ionization in the aurora dominates over photoionization at high latitudes. Indeed, Cassini radio occultations by Saturn point to a clear trend of increasing electron density with latitude (Kliore et al. 2009) that agrees with three-dimensional model calculations including auroral precipitation (Müller-Wodarg et al. 2012). In addition to meridional trends, there is evidence for diurnal variations, although this evidence is less clear (Kliore et al. 2009). The observed electron densities point to a surprising complexity in giant planet ionospheres that will continue to provide interesting problems for future studies. Such efforts would be greatly advanced by any observations of relative ion composition, which may in fact be obtained for the first time during the Cassini Grand Finale tour.

## Titan

Titan is Saturn's largest moon and the most characteristic example of a hazy environment in our solar system. Titan can serve as a reference for hazy exoplanets (Robinson et al. 2014); thus we focus here on the mechanisms responsible for the formation of hazes in this atmosphere, as revealed by the latest observations from the Cassini-Huygens mission. Titan is far too complex to be described in detail here, and interested readers are referred to recent reviews of this atmosphere (Müller-Wodarg et al. 2014).

### Photochemistry

Titan's atmosphere is dominantly composed of molecular nitrogen ($N_2$) with trace amounts of methane ($CH_4$) and carbon monoxide (CO) (Niemann et al. 2010; Yelle et al. 2008). Energy deposition in the upper atmosphere is driven mainly by high energy insolation; photons with $\lambda < 1000$ A are responsible for the photolysis of $N_2$ close to 1100 km, while Lyman-α photons break up $CH_4$ with a peak photolysis rate close to 800 km (Lavvas et al. 2011a). Titan does not have an intrinsic magnetic field, but as it orbits it is subjected to Saturn's variable magnetosphere. Energetic particles accelerated along the magnetic field lines are deposited in Titan's upper atmosphere and provide a secondary contribution to the $N_2$ destruction, at altitudes close to 1200 km (see Galand et al. 2014 for more details).

The primary products of $N_2$ and $CH_4$ photolysis provide the building blocks for the formation of larger molecules in Titan's atmosphere. For example, methyl radicals ($CH_3$) produced in the photolysis of methane can recombine to form ethane ($C_2H_6$) molecules, while excited nitrogen atoms ($N^2D$) formed in the photolysis of nitrogen react with methane leading eventually to the production of hydrogen cyanide (HCN) molecules. These are just the first steps of photochemistry in Titan's atmosphere, since the produced molecules are subsequently dissociated by solar radiation and the new products form other molecules. This mechanism allows for the formation of perpetually larger structures in Titan's atmosphere and terminates with the formation of photochemical aerosols, i.e., hazes.

The above "schematic" picture of atmospheric photochemistry can be separated into two different modes: the neutral and the ion contribution. Neutral chemistry driven mainly by Ly-α and lower energy photons is active in the whole atmosphere and is responsible for the bulk of the main photochemical products observed (see review by Vuitton et al. 2014). Ion chemistry, although limited to the ionosphere, is characterized by much faster reaction rates than the neutral chemistry, while it also allows for chemical pathways that are not possible through neutral reactions (Vuitton et al. 2007, 2008). These two characteristics lead to the rapid formation of macromolecules in Titan's thermosphere (Waite et al. 2007). The role of ion chemistry became apparent when the mass spectrometers of the Cassini orbiter discovered more than 50 positive ions in the mass range between 1 and 100 Dalton/charge (Da/q) (Vuitton et al. 2007), while the picture was further completed with the detection of multiple negative ions in the same mass range (Coates et al. 2007). Detailed chemical networks are required to identify the intricate pathways leading to the formation of these species, and state-of-the-art models are able to reproduce the composition constraints from the observations of both neutral and ion species (Vuitton et al. 2014).

## Aerosol Formation

Cassini observations had more surprises to reveal. Measurements at larger masses show an even more significant population of positive and negative ions (Fig.7). At the deepest altitudes probed (∼900 km), positive ions grow up to a few hundred Da/q (Crary et al. 2009), while negative ions masses up to 10,000 Da/q were detected (Coates et al. 2007). Such large molecules are unprecedented in planetary thermospheres and are a clear demonstration of efficient molecular growth taking

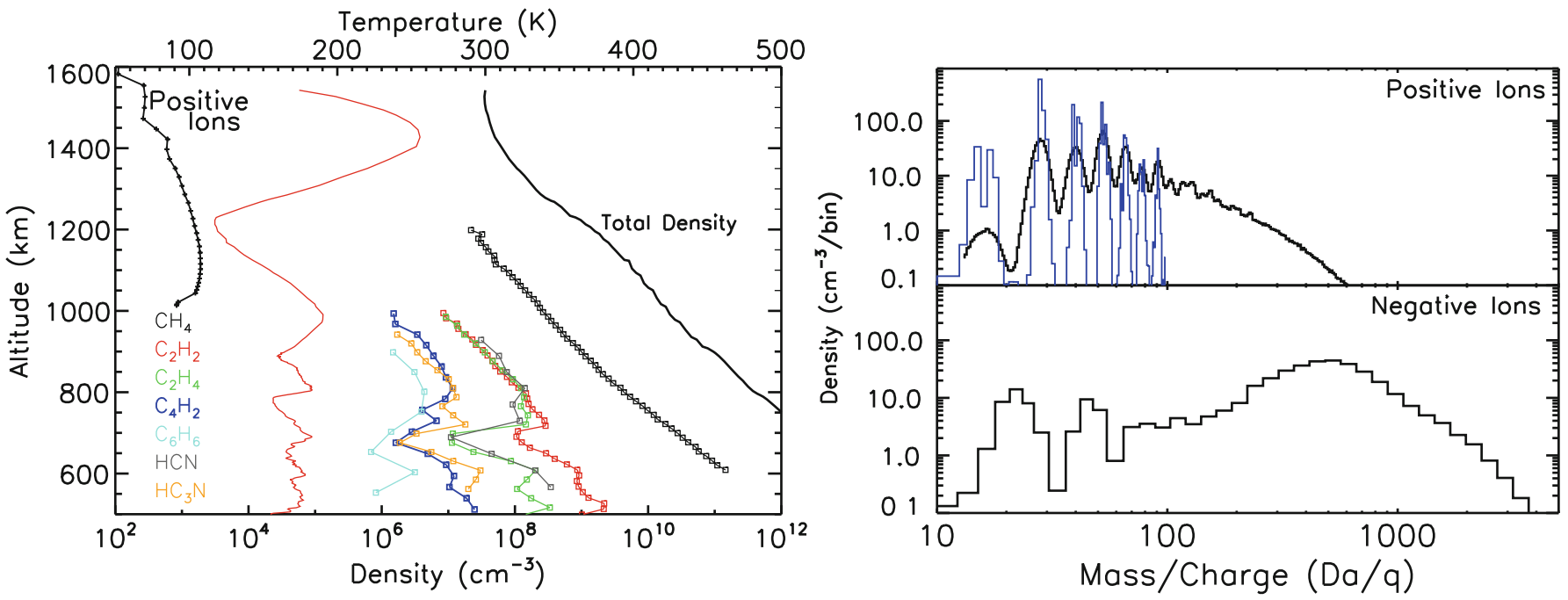

**Fig. 7** *Left*: density profiles of different components of Titan's atmosphere retrieved from Cassini-Huygens observations. The thick solid *black* and *red* lines present the total density (dominated by $N_2$) and temperature measured by Huygens/HASI (Fulchignoni et al. 2005). Colored lines with squares present retrievals of $CH_4$ and other neutral components from Cassini/UVIS occultations (Koskinen et al. 2011), while the *thin black* line with crosses presents the total positive ion density from Cassini/INMS measurements (Vuitton et al. 2009). *Right*: mass spectra of positive and negative ions observed by Cassini mass spectrometers (Coates et al. 2007; Vuitton et al. 2009; Crary et al. 2009). The INMS observations (*blue line*) have a higher mass resolution relative to the CAPS/IBS measurements (*black line*, *positive ions*) but are limited up to 100 Da (see Lavvas et al. 2013 for more details)

place in this atmosphere. Theoretical studies for the formation of these large ions demonstrate that they are the first steps of aerosol formation (Lavvas et al. 2013).

Yet, aerosol growth does not terminate in the ionosphere. The aerosol mass flux produced in the atmosphere is only a tenth of the flux observed in the lower atmosphere (Wahlund et al. 2009). Further growth of the particles' bulk mass is only possible through neutral chemical reactions on their surface (heterogeneous chemistry). The large abundance of radicals generated from the photochemistry in the upper atmosphere, along with the extended residence time of the particles in the atmosphere due to Titan's low gravity, outbalances the lower reaction rates of neutral relative to ion reaction rates and allows for an efficient increase of the aerosol mass flux below the ionosphere (Lavvas et al. 2011b). Thus, both ion and neutral chemistry contributions are important for the formation of aerosols in Titan's atmosphere, the first for initiating the aerosol formation and the second for defining their final mass flux.

## Energetics

Heating by solar EUV radiation and energetic particles from Saturn's magnetosphere and radiative cooling by HCN emissions are the dominant factors controlling the thermal structure of Titan's upper atmosphere (Yelle et al. 1991). These two main processes generate an average temperature of 150 K, which is consistent with the Cassini observations (Snowden and Yelle 2014). However, the observations also reveal significant altitude structure in the temperature profile of the upper atmosphere (Fulchignoni et al. 2005), as well as a strong temporal variability of the order of 60 K (Snowden et al. 2013). Theoretical studies demonstrate that this variability is not related to the temporally variable energy input from Saturn's magnetosphere, but could be affected by the dissipation of waves in this part of the atmosphere (Snowden and Yelle 2014; Cui et al. 2014).

Aerosols can interact strongly with the radiation field, therefore affecting the thermal structure of the atmosphere. This is well established for Titan's lower atmosphere where the particles have an effective size of the order of 2–3 μm and are responsible for heating the stratosphere and cooling the surface (see West et al. 2014 and references therein). Occultations of Titan's atmosphere at UV wavelengths reveal the presence of hazes in the upper atmosphere as well (Liang et al. 2007; Koskinen et al. 2011), verifying that indeed aerosol formation starts there. However, the role of these nascent aerosol particles in the thermal structure of the upper atmosphere is still under investigation. At those altitudes aerosol particles have smaller sizes, but higher populations than in the lower atmosphere, while their optical properties are unknown. Further analyses of Cassini observations and modeling are required to decipher the optical properties of the aerosols and their role in the energy balance of the Titan's upper atmosphere.

## References

Badman SV, Branduardi-Raymont G, Galand M et al (2015) Auroral processes at the giant planets: energy deposition, emission mechanisms, morphology and spectra. Space Sci Rev 187:99–179. doi:10.1007/s11214-014-0042-x

Ben-Jaffel L, Prange R, Sandel BR et al (1995) New analysis of the Voyager UVS H Lyman-α emission of Saturn. Icarus 113:91–102. doi:10.1006/icar.1995.1007

Bertaux JL, Leblanc F, Witasse O, Quemerais E, Lilensten J et al (2005) Discovery of an aurora on Mars. Nature 435:790–794. doi:10.1038/nature03603

Bhardwaj A, Thampi SV, Das TP et al (2016) On the evening time exosphere of Mars: result from MENCA aboard Mars orbiter mission. GRL 43:1862–1867. doi:10.1002/2016GL067707

Bougher SW, Cravens TE, Grebowsky J, Luhmann J (2015a) The aeronomy of Mars: characterization by MAVEN of the upper atmosphere reservoir that regulates volatile escape. Space Sci Rev 195:423–456. doi:10.1007/s11214-014-0053-7

Bougher SW, Jakosky B, Halekas J, Grebowsky J, Luhmann J (2015b) Early MAVEN deep dip campaign reveals thermosphere and ionosphere variability. Science 350:0459. doi:10.1126/science.aad0459

Catling DC, Zahnle KJ (2009) The planetary air lead. Sci Am 300:36–43. doi:10.1038/scientificamerican0509-36

Choudhary RK, Ambili KM, Choudhury S, Dhanya MB, Bhardwaj A (2016) On the origin of the ionosphere at the moon using results from Chandrayaan-1 S band radio occultation experiment and a photochemical model. Geophys Res Lett 43:10025–10033. doi:10.1002/2016GL070612

Clancy RT, Sandor BJ, García Muñoz A, Lefèvre F, Smith MD, Wolff MJ, Montmessin F, Murchie SL, Nair H (2013) First detection of Mars atmospheric hydroxyl: CRISM near-IR measurement versus LMD GCM simulation of OH Meinel band emission in the Mars polar winter atmosphere. Icarus 226:272–281. doi:10.1016/j.icarus.2013.05.035

Clancy RT, Sandor BJ, Hoge J (2015) Doppler winds mapped around the lower thermospheric terminator of Venus: 2012 solar transit observations from the James Clerk Maxwell Telescope. Icarus 254:233–258. doi:10.1016/j.icarus.2015.03.031

Coates AJ, Crary FJ, Lewis GR et al (2007) Discovery of heavy negative ions in Titan's ionosphere. GRL 34:L22103. doi:10.1029/2007GRL030978

Connerney JEP, Waite JH (1984) New model of Saturn's ionosphere with an influx of water from the rings. Nature 312:136–138. doi:10.1038/312136a0

Crary FJ, Magee BA, Mandt K et al (2009) Heavy ions, temperatures and winds in Titan's ionosphere: combined Cassini CAPS and INMS observations. PSS 57:1847–1856. doi:10.1016/j.pss.2009.09.006

Cravens TE, Brace LH, Taylor HA Jr, Quenon SJ, Russell CT et al (1982) Disappearing ionospheres on the nightside of Venus. Icarus 51:271–282. doi:10.1016/0019-1035(82)90083-5

Cui J, Yelle RV, Li T et al (2014) Density waves in Titan's upper atmosphere. J Geophys R 119:490–518. doi:10.1002/2013JA019113

Dennerl K (2008) X-rays from Venus observed with Chandra. Planet Space Sci 56:1414–1423. doi:10.1016/j.pss.2008.03.008

Drossart P, Maillard J-P, Caldwell J et al (1989) Detection of $H_3^+$ on Jupiter. Nature 340:539–541. doi:10.1038/340539a0

Drossart P, Fouchet Th, Crovisier J et al (1999) Fluorescence in the 3 micron bands of methane on Jupiter and Saturn from ISO/SWS observations. In: Cox P, Kessler MF (eds) The Universe as seen by ISO. ESA special publication, vol 427. pp 169–172

Edgington SG, Spilker LJ (2016) Cassini's grand finale. Nat Geosci 9:472–473

Feuchtgruber H, Lellouch E, de Graauw T et al (1997) External supply of oxygen to the atmospheres of the giant planets. Nature 389:159–162. doi:10.1038/38236

Fletcher LN, Orton GS, Teanby NA, Irwin PGJ, Bjoraker GL (2009) Methane and its isotopologues on Saturn from Cassini/CIRS observations. Icarus 199:351–367. doi:10.1016/j.icarus.2008.09.019

Forget F, Montmessin F, Bertaux J-L, González-Galindo F, Lebonnois S et al (2009) Density and temperatures of the upper Martian atmosphere measured by stellar occultations with Mars Express SPICAM. JGR:114, E01004. doi:10.1029/2008JE003086

Fox JL (2009) Morphology of the dayside ionosphere of Mars: implications for ion outflows. JGR 114:E12005. doi:10.1029/2009JE003432

Fox JL, Bougher SW (1991) Structure, luminosity and dynamics of the Venus thermosphere. Space Sci Rev 55:357–489. doi:10.1007/BF00177141

Fox JL, Sung KY (2001) Solar activity variations of the Venus thermosphere/ionosphere. J Geophys Res 106:21305–21335. doi:10.1029/2001JA000069

Friedson AJ, Wong A-S, Yung YL (2002) Models for polar haze formation in Jupiter's stratosphere. Icarus 158:389–400. doi:10.1006/icar.2002.6885

Fulchignoni M, Ferri F, Angrilli F et al (2005) In situ measurements of the physical characteristics of Titan's environment. Nature 438:785–791. doi:10.1038/nature043114

Galand M, Coates AJ, Cravens TE, Wahlund J-E (2014) Titan's ionosphere. In: Titan. Interior, surface, atmosphere, and space environment. Cambridge Planetary Science. Cambridge University Press, Cambridge, UK ISBN 9780521199926

García Muñoz A, Mills FP, Piccioni G, Drossart P (2009) The near-infrared nitric oxide nightglow in the upper atmosphere of Venus. Proc Natl Acad Sci 106:985–988. doi:10.1073/pnas.0808091106

González-Galindo F, López-Valverde MA, Forget F, García-Comas M, Millour E, Montabone L (2015) Variability of the Martian thermosphere during eight Martian years as simulated by a ground-to-exosphere global circulation model. JGR Planet 120:2020–2035. doi:10.1002/2015JE004925

Greathouse TK, Gladstone GR, Moses JI et al (2010) New horizons Alice ultraviolet observations of a stellar occultation by Jupiter's atmosphere. Icarus 208:293–305. doi:10.1016/j.icarus.2010.02.002

Grebowsky JM, Benna M, Planet JMC, Collinson GA, Mahaffy PR, Jakosky BM (2017) Unique, non-earthlike, meteoritic ion behavior in upper atmosphere of Mars. Geophys Res Lett. doi:10.1002/2017GL072635

Grodent D (2015) A brief review of ultraviolet auroral emissions on giant planets. Space Sci Rev 187:23–50. doi:10.1007/s11214-014-0052-8

Herbert F, Sandel BR (1999) Ultraviolet observations of Uranus and Neptune. Planet Space Sci 47:1119–1139. doi:10.1016/S0032-0633(98)00142-1

Hinson DP, Flasar FM, Kliore AJ, Schinder PJ, Twicken JD, Herrera RG (1997) Jupiter's ionosphere: results from the first Galileo radio occultation experiment. Geophys Res Lett 24:2107–2110. doi:10.1029/97GL01608

Huestis DL, Slanger TG, Sharpee BD, Fox JL (2010) Chemical origins of the Mars ultraviolet dayglow. Faraday Discuss 147:307–322. doi:10.1039/C003456H

Hunten DM (1993) Atmospheric evolution of the terrestrial planets. Science 259:915–920. doi:10.1126/science.259.5097.915

Jakosky BM, Grebowsky JM, Luhmann JG, Connerney J, Eparvier F et al (2015) MAVEN observations of the response of Mars to an interplanetary coronal mass ejection. Science 350:0210. doi:10.1126/science.aad0210

Jakosky BM, Slipski M, Benna M, Mahaffy P, Elrod M, Yelle R, Stone S, Alsaeed N (2017) Mars' atmospheric history derived from upper-atmosphere measurements of $^{38}$Ar/3$^{6}$Ar. Science 355:1408–1410. doi:10.1126/science.aai7721

Keating GM, Bertaux JL, Bougher SW, Dickinson RE, Cravens TE, Hedin AE (1985) Models of Venus neutral upper atmosphere – structure and composition. Adv Space Res 5:117–171. doi:10.1016/0273-1177(85)90200-5

Kim YH, Fox JL (1994) The chemistry of hydrocarbon ions in the Jovian ionosphere. Icarus 112:310–325. doi:10.1006/icar.1994.1186

Kim YH, Fox JL, Black JH, Moses JI (2014) Hydrocarbon ions in the lower ionosphere of Saturn. J Geophys Res 119:384–395. doi:10.1002/2013JA019022

Kliore AJ, Patel IR, Lindal GF et al (1980) Structure of the ionosphere and atmosphere of Saturn from Pioneer 11 Saturn radio occultation. J Geophys Res 85:5857–5870. doi:10.1029/JA085iA11p05857

Kliore AJ, Nagy AF, Marouf EA et al (2009) Midlatitude and high-latitude electron density profiles in the ionosphere of Saturn obtained by Cassini radio occultation observations. J Geophys Res 114:A04315. doi:10.1029/2008JA013900

Koskinen T, Yelle RV, Snowden D et al (2011) The mesosphere and thermosphere of Titan revealed by Cassini/UVIS stellar occultations. Icarus 216:507–534. doi:10.1016/j.icarus.2011.09.022

Koskinen TT, Sandel BR, Yelle RV et al (2013) The density and temperature structure near the exobase of Saturn from Cassini UVIS solar occultations. Icarus 226:1318–1330. doi:10.1016/j.icarus.2013.07.037

Koskinen TT, Sandel BR, Yelle RV et al (2015) Saturn's variable thermosphere from Cassini/UVIS occultations. Icarus 260:174–189. doi:10.1016/j.icarus.2015.07.008

Koskinen TT, Moses JI, West RA, Guerlet S, Jouchoux A (2016) The detection of benzene in Saturn's upper atmosphere. Geophys Res Lett 43:7895–7901. doi:10.1002/2016GL070000

Krasnopolsky VA (2002) Mars' upper atmosphere and ionosphere at low, medium, and high solar activities: implications for evolution of water. J Geophys Res (Planets) 107:5128. doi:10.1029/2001JE001809

Lammer H, Kasting JF, Chassefière E, Johnson RE, Kulikov YN, Tiang F (2008) Atmospheric escape and evolution of terrestrial planets and satellites. Space Sci Rev 139(1–4):399–436. doi:10.1007/s11214-008-9413-5

Lavvas P, Galand M, Yelle RV et al (2011a) Energy deposition and primary chemical products in Titan's upper atmosphere. Icarus 213:233–251. doi:10.1016/j.icarus.2011.03.001

Lavvas P, Sander M, Kraft M et al (2011b) Surface chemistry and particle shape: processes for the evolution of aerosols in Titan's atmosphere. Astrophys J 728:80. doi:10.1088/0004-637X/728/2/80. (11p)

Lavvas P, Yelle RV, Koskinen T et al (2013) Aerosol growth in Titan's ionosphere. PNAS 110:2729–2734. doi:10.1073/pnas.1217059110

Lellouch E, Moreno R, Orton GS et al (2015) New constraints on the CH4 vertical profile in Uranus and Neptune from Herschel observations. A&A 579:A121. doi:10.1051/0004-6361/201526518

Liang MC, Yung YL, Shemansky DE (2007) Photolytically generated aerosols in the mesosphere and thermosphere of Titan. Astrophys J 661:199–202. doi:10.1086/518785

Lindal GF (1992) The atmosphere of Neptune: an analysis of radio occultation data acquired with Voyager 2. AJ 103:967–982. doi:10.1086/116119

Lindal GF, Sweetnam DN, Eshleman VR (1985) The atmosphere of Saturn: an analysis of the Voyager radio occultation measurements. ApJ 90:1136–1146. doi:10.1086/113820

Lindal GF, Lyons JR, Sweetnam DN, Eshleman VR, Hinson DP, Tyler GL (1987) The atmosphere of Uranus: results of radio occultation measurements with Voyager 2. J Geophys Res 92:14987–15001. doi:10.1029/JA092iA13p14987

Liu W, Dalgarno A (1996) The ultraviolet spectra of the Jovian dayglow. ApJ 462:502–518. doi:10.1086/177168

Mahieux A, Vandaele AC, Robert S, Wilquet V, Drummond R et al (2015) Rotational temperatures of Venus upper atmosphere as measured by SOIR on board Venus Express. Planet Space Sci 113:347–358. doi:10.1016/j.pss.2014.12.020

Majeed T, McConnell JC, Gladstone GR (1999) A model analysis of Galileo electron densities on Jupiter. Geophys Res Lett 26:2335–2338. doi:10.1029/1999GL900530

Majeed T, Waite JH Jr, Bougher SW et al (2005) Process of equatorial thermal structure at Jupiter: an analysis of the Galileo temperature profile with a three-dimensional model. J Geophys Res 110:E12007. doi:10.1029/2004JE002351

Mangold N, Baratoux D, Witasse O, Encrenaz T, Sotin C (2016) Mars: a small terrestrial planet. Astron Astrophys Rev 24:15. doi:10.1007/s00159-016-0099-5

Matcheva KI, Strobel DF, Flasar FM (2001) Interaction of gravity waves with ionospheric plasma: implications for Jupiter's ionosphere. Icarus 152:347–365. doi:10.1006/icar.2001.6631

Meier RR (1991) Ultraviolet spectroscopy and remote sensing of the upper atmosphere. Space Sci Rev 58:1–185. doi:10.1007/BF01206000

Mendillo M, Nagy A, Waite JH (eds) (2002) Atmospheres in the solar system. American Geophysical Union, Geophysical Monograph, Washington, DC, p 130

Miller KL, Knudsen WC, Spenner K, Whitten RC, Novak V (1980) Solar zenith angle dependence of ionospheric ion and electron temperatures and density on Venus. J Geophys Res Space Phys 85:7759–7764. doi:10.1029/JA085iA13p07759

Miller S, Stallard T, Smith C et al (2006) ${H_3}^+$: the driver of giant planet atmospheres. Phil Trans R Soc A 364:3121–3137. doi:10.1098/rsta.2006.1877

Miller S, Stallard T, Melin H, Tennyson J (2010) ${H_3}^+$ cooling in planetary atmospheres. Faraday Discuss 147:283–291. doi:10.1039/c004152c

Moses JI, Bass SF (2000) The effects of external material on the chemistry and structure of Saturn's ionosphere. J Geophys Res 105:7013–7052. doi:10.1029/1999JE001172

Moses JI, Rages K, Pollack JB (1995) An analysis of Neptune's stratospheric haze using high phase-angle Voyager images. Icarus 113:232–266. doi:10.1006/icar.1995.1022

Mueller-Wodarg ICF, Strobel DF, Moses JI, Waie JH, Crovisier J, Yelle RV, Bougher SW, Roble RG (2008) Neutral atmospheres. Space Sci Rev 139:191–234. doi:10.1007/s11214-008-9404-6

Müller-Wodarg ICF, Mendillo M, Yelle RV et al (2006) A global circulation model of Saturn's thermosphere. Icarus 180:147–160. doi:10.1016/j.icarus.2005.09.002

Müller-Wodarg ICF, Moore L, Galand M, Miller S, Mendillo M (2012) Magnetosphere-atmosphere coupling at Saturn: 1 – response of thermosphere and ionosphere to steady state polar forcing. Icarus 221:481–494. doi:10.1016/j.icarus.2012.08.034

Müller-Wodarg I, Griffith CA, Lellouch E, Cravens TE (2014) Titan. Interior, surface, atmosphere, and space environment. Cambridge Planetary Science. Cambridge University Press, Cambridge, UK ISBN 9780521199926

Müller-Wodarg ICF, Bruinsma S, Marty J-C, Svedhem H (2016) In situ observations of waves in Venus's polar lower thermosphere with Venus Express aerobraking. Nat Phys 12:767–771. doi:10.1038/nphys3733

Nagy AF, Balogh A, Cravens TE, Mendillo M, Mueller-Wodarg I (eds) (2008) Comparative

Niemann HB, Atreya SK, Demick JE et al (2010) Composition of Titan's lower atmosphere and simple surface volatiles as measured by the Cassini-Huygens probe gas chromatograph mass spectrometer experiment. J Geophys Res 115(E):12006. doi:10.1029/2010JE003659

O'Donoghue J, Stallard TS, Melin H et al (2013) The domination of Saturn's low-latitude ionosphere by ring rain. Nature 496:193–195. doi:10.1038/nature12049

Parkinson CD, Griffioen E, McConnell JC, Gladstone GR, Sandel BR (1998) He 584 Å dayglow at Saturn: a reassessment. Icarus 133:210–220. doi:10.1006/icar.1998.5926

Pätzold A, Häusler B, Bird MK, Tellmann S, Mattei R, Asmar SW, Dehant V, Eidel W, Imamura T, Simpson RA, Tyler GL (2007) The structure of Venus' middle atmosphere and ionosphere. Nature 450:657–660. doi:10.1038/nature06239

Rees MH (1989) Physics and chemistry of the upper atmosphere. Cambridge University Press, London

Robinson TD, Maltagliati L, Marley M et al (2014) Titan solar occultation observations reveal transit spectra of a hazy world. PNAS 111:9042–9047. doi:10.1073/pnas.1403473111

Sánchez-Lavega A, García Muñoz A, García-Melendo E, Pérez-Hoyos S, Gómez-Forrelad et al (2015) An extremely high-altitude plume seen at Mars' morning terminator. Nature 518:525–528. doi:10.1038/nature14162

Schneider NM, Deighan JI, Jain SK, Stiepen A, Stewart AIF et al (2015) Discovery of diffuse aurora on Mars. Science 350:0313. doi:10.1126/science.aad0313

Schubert GMP, Hickey MP, Walterscheid RL (2003) Heating of Jupiter's thermosphere by the dissipation of upward propagating acoustic waves. Icarus 163:398–413. doi:10.1016/S0019-1035(03)00078-2

Schunk RW, Nagy AF (2000) Ionospheres. Physics, plasma physics, and chemistry. Cambridge University Press, Cambridge

Seiff A, Kirk DB, Knight TCD et al (1998) Thermal structure of Jupiter's atmosphere in the North Equatorial Belt, near the edge of a 5 μm hot spot. J Geophys Res 103:22857–22890. doi:10.1029/98JE01766

Slanger TG, Cosby PC, Huestis DL, Bida TA (2001) Discovery of the atomic oxygen green line in the Venus night airglow. Science 291:463–465. doi:10.1126/science.291.5503.463

Smith CGA, Aylward AD (2009) Coupled rotational dynamics of Jupiter's thermosphere and ionosphere. Ann Geophys 27:199–230. doi:10.5194/angeo-27-199-2009

Smith CGA, Aylward AD, Millward GH et al (2007) An unexpected cooling effect in Saturn's upper atmosphere. Nature 445:399–401. doi:10.1038/nature05518

Snowden D, Yelle RV (2014) The thermal structure of Titan's upper atmosphere, II: energetics. Icarus 228:64–77. doi:10.1016/j.icarus.2013.08.027

Snowden D, Yelle RV, Cui J et al (2013) The thermal structure of Titan's upper atmosphere, I: temperature profiles from Cassini INMS observations. Icarus 226:552–582. doi:10.1016/j.icarus.2013.06.006

Stevens MH, Strobel DF, Herbert F (1993) An analysis of the Voyager 2 ultraviolet spectrometer occultation data at Uranus: inferring heat sources and model atmospheres. Icarus 100:45–63. doi:10.1006/icar.1993.1005

Strobel DF, Koskinen T, Müller-Wodarg ICF (2016) Saturn's variable thermosphere. Astro-ph arXiv:1610.07669v1

Stubbs TJ, Glenar DA, Farrell WM, Vondrak RR, Collier MR et al (2011) On the role of dust in the lunar ionosphere. Planet Space Sci 59:1659–1664. doi:10.1016/j.pss.2011.05.011

Vervack RJ Jr, Moses JI (2015) Saturn's upper atmosphere during the Voyager era: reanalysis and modeling of the UVS occultations. Icarus 258:135–163. doi:10.1016/j.icarus.2015.06.007

Vogt MF, Withers P, Mahaffy PR, Benna M, Elrod MK et al (2015) Ionopause-like density gradients in the Martian ionosphere: a first look with MAVEN. Geophys Res Lett 42:8885–8893. doi:10.1002/2015GL065269

Volkov AN, Johnson RE, Tucker OJ, Erwin JT (2011) Thermally driven atmospheric escape: transition from hydrodynamic to Jeans escape. Astrophys J 729:L24. doi:10.1088/2041-8205/729/2/L24

Vuitton V, Yelle RV, McEwan MJ (2007) Ion chemistry and N-containing molecules in Titan's upper atmosphere. Icarus 191:722–742. doi:10.1016/j.icarus.2007.06.023

Vuitton V, Yelle RV, Cui J (2008) Formation and distribution of benzene on Titan. J Geophys Res 113(E):05007. doi:10.1029/2007JE002997

Vuitton V, Lavvas P, Yelle RV et al (2009) Negative ion chemistry in Titan's upper atmosphere. PSS 57:1558–1572. doi:10.1016/j.pss.2009.04.004

Vuitton V, Dutuit O, Smith MA et al (2014) Chemistry of Titan's atmosphere in Titan. Interior, surface, atmosphere, and space environment. Cambridge Planetary Science. Cambridge University Press, Cambridge, UK ISBN 9780521199926

Wahlund J-E, Galand M, Muller-Wodard I et al (2009) On the amount of heavy molecular ions in Titan's ionosphere. PSS 57:1857–1865. doi:10.1016/j.pss.2009.07.014

Waite JH Jr, Young DT, Cravens TE et al (2007) The process of Tholin formation in Titan's upper atmosphere. Science 316:870–875. doi:10.1126/science.1139727

West RA, Lavvas P, Anderson C et al (2014) Titan's Haze. In: Titan. Interior, surface, atmosphere, and space environment. Cambridge Planetary Science. Cambridge University Press, Cambridge, UK ISBN 9780521199926

Withers P (2012) How do meteoroids affect Venus's and Mars's ionospheres? EOS Trans Am Geophys Union 93:337–338. doi:10.1029/2012EO350002

Withers P, Pratt R (2013) An observational study of the response of the upper atmosphere of Mars to lower atmospheric dust storms. Icarus 225:378–389. doi:10.1016/j.icarus.2013.02.032

Withers P, Fillingim MO, Lillis RJ, Häusler B, Hinson DP et al (2012) Observations of the nightside ionosphere of Mars by the Mars Express radio Science experiment (MaRS). J Geophys Res Space Phys 117:A12307. doi:10.1029/2012JA018185

Wong A-S, Yung YL, Friedson AJ (2003) Benzene and haze formation in the polar atmosphere of Jupiter. Geophys Res Lett 30:1447. doi:10.1029/2002GL016661

Yelle RV (1991) Non-LTE models of Titan's upper atmosphere. Astrophys J 383:380–400. doi:10.1086/170796

Yelle RV, Miller S (2004) In: Bagenal F, Dowling TE, McKinnon WB (eds) Jupiter's thermosphere and ionosphere. Cambridge University Press, Cambridge, pp 185–218

Yelle RV, Herbert F, Sandel BR, Vervack RJ Jr, Wentzel TM (1993) The distribution of hydrocarbons in Neptune's upper atmosphere. Icarus 104:38–59. doi:10.1006/icar.1993.1081

Yelle RV, Griffith CA, Young LA (2001) Structure of Jovian stratosphere at the Galileo probe entry site. Icarus 152:331–346. doi:10.1006/icar.2001.6640

Yelle RV, Cui J, Muller-Wodarg ICF (2008) Methane escape from Titan's atmosphere. J Geophys Res 113(E):10003. doi:10.1029/2007JE003031

Young LA, Yelle RV, Young RE et al (1997) Gravity waves in Jupiter's thermosphere. Science 276:108–111. doi:10.1126/science.276.5309.108

Zhang TL, Lu QM, Baumjohann W et al (2012) Magnetic reconnection in the near Venusian magnetotail. Science 336:567–570. doi:10.1126/science.1217013

Zurbuchen TH, Raines JM, Slavin JA, Gershman DJ, Gilbert JA et al (2011) MESSENGER observations of the spatial distribution of planetary ions near Mercury. Science 333:1862–1865. doi:10.1126/science.1211302